\documentclass[dvipdfmx]{ws-ijgmmp}
\usepackage[scr=rsfs]{mathalpha}
\usepackage{xcolor}
\usepackage{tikz}
\usepackage{titleps}
\usepackage{enumerate}
\usepackage{enumitem}

\begin{document}

\markboth{H. Sakuma and et al.}
{Reexamination of the hierarchy problem}

%
\catchline{}{}{}{}{}
%

\title{REEXAMINATION OF THE HIERARCHY PROBLEM \\ FROM A NOVEL GEOMETRIC PERSPECTIVE
}
\author{HIROFUMI SAKUMA$^{\dag}$, IZUMI OJIMA$^{\S}$}


\address{Research Origin for Dressed Photon, \\ 
Yokohama City, Kanagawa 221-0022, Japan\,\footnote{Bldg.1-1F, 3-13-19 Moriya-cho,
Kanagawa-ku, Yokohama, Kanagawa, 221-0022 Japan}\\
\email{$^{\dag}$sakuma@rodrep.or.jp, $^{\S}$ojima@gaia.eonet.ne.jp}}



\author{KAZUYA OKAMURA}
\address{College of Engineering \\
Chubu University Center for Mathematical Science and Artificial Intelligence \\ 
1200 Matsumoto-cho, Kasugai-shi, Aichi 487-8501, Japan\\
k.okamura.renormalizable@gmail.com}

\maketitle

\begin{history}
\received{(Day Month Year)}
\revised{(Day Month Year)}
\end{history}

\begin{abstract}
A lucid interpretation of the longstanding hierarchy problem is proposed based on the unconventional model of the universe recently proposed by the authors. Our heuristic cosmological model is developed by considering Penrose and Petit's original ideas as the Weyl curvature hypothesis, conformal cyclic cosmology, and the twin universe model. The uniqueness of our model lies in its incorporation of \textit{dark energy and matter}, and its single key parameter, adjusted by observational data, is the radius ($R_{dS}$) of a four-dimensional ($4D$) hypersphere called de Sitter space. We presuppose that our $4D$ universe originated from the spontaneous conformal symmetry (SCS) breaking of a light field with a null distance. We show that in this SCS breaking state, the energy--momentum tensor of the space-like electromagnetic field, whose existence is inevitable for quantum electromagnetic field interactions (Greenberg--Robinson theorem), becomes isomorphic to the divergence-free Einstein tensor in the general theory of relativity. Furthermore, we reveal the $R_{dS}$ dependency of the $4D$ gravitational field. Based on these findings, we show an intriguing relation between the magnitude of the gravitational coupling constant and $R_{dS}$. A solution to the hierarchy problem is derived by assuming that $R_{dS}$ depends on the \lq\lq newly defined cosmological time\rq\rq.
\end{abstract}


\section{Introduction}
This study aims to shed new light on the unresolved hierarchy problem in particle physics. The problem stems from the incredible difference existing between, for instance, the magnitudes of electromagnetic and gravitational coupling constants. This inspired the investigations of Dirac\cite{dirac1, dirac2}, Milne\cite{milne}, and other researchers\rq $\;$\cite{canuto, fran, reev, peeb}, wherein they refute the hypothesis on the invariance property of physical constants, particularly of gravitational constant $G$. To understand the problem, the possibility of simultaneous temporal change of all physical constants was examined first by Troitskii\cite{troit} and explored further by Petit\cite{petit1, petit2}. At the end of the 20th century, an entirely different approach to explaining the hierarchy problem was presented by Randoll and Sundrum\cite{rand}, who introduced a notion of warped extra dimensions based on the brane dynamics in superstring theory. Notably, our model is distinct from the abovementioned two in the sense that it incorporates dark energy and matter as crucial components. However, our model inherits important characteristics from those preceding researches. For instance, it addresses the \lq\lq temporal\rq\rq evolution of the magnitude of the cosmological constant associated with the extra expansion of the universe in the fifth dimension ($D$) perpendicular to our $4D$ universe.

The elucidation of new ideas is frequently accompanied by the dissemination of important knowledge. Although such knowledge might be new or old, it is often not widespread in the relevant community. In our case, we have three pieces of such important knowledge: two of which, the Greenberg--Robinson (GR) theorem\cite{jost, antonio} and the micro--macro duality (MMD) theory of Ojima\cite{ojima1, ojima2}, are considered old information, whereas the Clebsch dual space-like electromagnetic (CDSE) theory\cite{sakuma0, sakuma1, sakuma2} built on the two preceding pieces of knowledge is considered new knowledge. As demonstrated by the Schr\"{o}dinger\rq s cat, the prevailing understanding of the invisible microscopic quantum world and its connection to the visible macroscopic world is unsatisfactory. The MMD theory, built on the basic philosophy of the Araki--Haag--Kastler formulation of the algebraic quantum field theory (QFT), is an ambitious scheme aiming to provide a solution to the Schr\"{o}dinger\rq s cat problem in quantum physics. Notably, herein, the term \lq\lq quantum physics'' denotes relativistic quantum physics unless otherwise stated.

In the field of quantum physics, the prevailing knowledge is quantum mechanics (QM) with finite degrees of freedom. The knowledge required for this study is not that on QM but on QFT with \textit{infinite degrees of freedom}, whose dynamical behaviors are considerably more complex than those of QM. Moreover, the crucial difference between QM and QFT lies in the number of sectors, where a sector is defined as the dynamical domain in which the principle of superposition on the quantum states holds good. For a QM system, only one sector exists (Stone--von Neumann theorem\cite{mack}), whereas for a system described by QFT, multiple unitarily inequivalent sectors exist (initially highlighted by van Hove in 1952). Initially, the abovementioned van Hove phenomena were considered as the pathological characteristics of a QFT-based system with infinite degrees of freedom. However, later, the existence of such multiple sectors was determined to be the source of dynamical richness with which the visible macroscopic world can emerge from the invisible quantum microscopic world.

Ojima\rq s MMD theory was developed from the Doplicher--Haag--Roberts (DHR)\rq s original sector theory\cite{dhr1, dr1}, which attempted to formulate the QFT based solely on the maximal set of observables. In addition, Ojima developed the theory further by generalizing it in such a way that it could cope with the important physical mechanism of (spontaneous) symmetry breaking of given quantum fields. Notably, in the MMD theory, the original sector notion based on irreducible representations is revised into a generalized one with factor representation. In the revised variant, the factor representation centers feature attributes of either trivial or nontrivial, corresponding to microquantum or macroclassical factor representations, respectively. In addition, these generalized sectors are mutually disjoint (separated by the absence of intertwiners), which is a stronger notion than unitary inequivalence. Thus, the MMD theory became a comprehensive QFT, which can consistently connect the quantum world, where the algebra of physical quantities obeys the noncommutative law, to the classical world, where the corresponding algebra is replaced by the one with commutative law. Thus, the \textit{MMD theory embodies the mathematically rigorous quantum--classical correspondence.} A plain explanation of the essence of the MMD theory without resorting to sophisticated mathematical representations has been provided by Sakuma et al.\cite{sakuma5}.

Clarifying the mechanism of nonlinear field interactions would pose the biggest dynamical challenge in QFT studies. As is well known, the essential features of Fock spaces, such as the positive energy spectra in the state vector space generated by repeated applications of the creation operators on the Fock vacuum $|0\rangle$ (under the cyclicity assumption), are derived from the following Eq. (\ref{eqn:1}):
\begin{equation}
\phi^{(f)}(x^{0},\tilde{x}) = \int \frac{d^{3}\tilde{k}}{\sqrt{(2\pi)^{3}2E_{k}}}[a(\tilde{k})\exp {(-ik_{\nu }x^{\nu })} + a^{\dagger }(\tilde{k})\exp {(ik_{\nu }x^{\nu })}],  \label{eqn:1}
\end{equation}
where ($a^{\dagger }(\tilde{k})$, $a(\tilde{l})$) and ($\tilde{x}$ and $\tilde{k}$), respectively, denote a pair of creation--annihilation operators and of three vectors comprising spatial components of $x^{\mu }$ and $k_{\nu}$, with $E_{k}$ defined by $E_{k}: = \sqrt{(\tilde{k})^{2} + (m_{0})^{2}}$. Although the $\phi^{f}$ field thus constructed embodies the wave--particle duality of a quantum system, the $\phi^{f}$ field and the associated 4-momentum $p^{\mu}$ are, respectively, restricted by the following linear Klein--Gordon (KG) equation and the on-shell condition of the form
\begin{equation}
[\hbar^{2}\partial^{\nu}\partial_{\nu} + (m_{0}c)^{2}]\phi^{(f)} = 0,\;\;\;\; p_{\nu}p^{\nu} = (m_{0}c)^{2}\geq 0,  \label{eqn:3} 
\end{equation}
where the sign convention of $(+,-,-,-)$ is used for the Lorentzian metric tensor. Regarding these expressions, however, with the linearity of Eq. (\ref{eqn:3}) overlooked, the abovementioned characteristic features of Fock spaces are frequently misinterpreted as the universal structure to be found in interacting multiparticle systems. Accordingly, $|0\rangle$ becomes as mysterious as the creation of everything from emptiness.

In the quest to clarify quantum field interactions, the GR theorem stands as a noteworthy forerunner. In fact, it provides a clear-cut mathematical criterion for distinguishing nonlinear quantum field interactions from the time evolution of the free modes ($\phi^{(f)}$) with the second quantization described in Eq. (\ref{eqn:1}). The GR theorem states that \textit{if the Fourier transform ($\varphi(p)$) of a given quantum field ($\varphi(x)$) does not contain an off-shell space-like momentum ($p_{\mu}$) satisfying $p_{\nu}p^{\nu} < 0$, then $\varphi(x)$ is a generalized free field.} We believe that this theorem highlights the missing factor in the presently incomplete QFT. In the appendix, we provide a sketchy overview of the theorem derived based on Wightman axioms. In the field of elementary particle physics, such off-shell quantities are conventionally dismissed as nonphysical ghosts or extremely short-lived transitory entities. Consequently, the wide physical science communities have not come to an important research conclusion on off-shell quantities in general, despite their importance.

In conventional quantum electrodynamics (QED), the longitudinal mode of a field is eliminated as a nonphysical mode, although, classically, the Coulomb mode plays an important role in electromagnetic field interactions. This well-known fact highlights the limitation of the prevailing QED theory in affording a satisfactory explanation of the consistent connection between the quantum world and classical one. As the first step to a solution, based on the GR theorem and MMD theory, the first author successfully extended a free electromagnetic field into a space-like momentum domain \cite{sakuma6}. As this extension was achieved by introducing Clebsch variables \cite{clebsch} used in the study of the Hamiltonian structure of barotropic fluid, we call the extended electromagnetic field the Clebsch dual (CD) field. Notably, energy--momentum tensor of the field assumes the same form as Einstein's tensor in the theory of general relativity. Namely, the CD field bears the characteristics of spacetime itself in the sense of general relativity and can be shown to be a promising candidate for understanding the enigmatic dark energy field \cite{sakuma1}. As the CD field plays an important role in our discussion of the hierarchy problem, we will recapitulate its necessary points in Section 2.

As stated at the beginning of this section, we aim to present a novel view of the hierarchy problem in Section 3. To do that, we think it is inevitable to explore the different ideas published recently in our series of papers\cite{sakuma1, sakuma3, sakuma4}. Accordingly, considering that majority of readers would be unfamiliar with those ideas, in Section 2, we provide a bird\rq s eye view of the grand design of our research. The overview encompasses all individual aspects necessary in the pursuit of what we call \lq\lq off-shell science\rq\rq, originating from dressed photon (DP) studies in the field of nanophotonics. Thereafter, in Section 4, we highlight an intriguing finding on the meaning of Maldacena duality\cite{mald} in superstring theory, which applies to the early stages of our new model of the universe. 
Finally, the summary and conclusions are presented in Section 5.

\section{Overview of our ongoing research on reformulating the quantum field theory}
As stated in the preceding section, one of the focal points of our research on the \textit{extended light field} is off-shell quantum fields involving quantum field interactions. The existence of off-shell (superluminal wave-) momenta implies, as in the case of quantum spin entanglement, an instantaneous connectedness of the entire universe. This naturally leads us to some cosmological problems that are not necessarily within the scope of this study. As cosmology remains a vast, unknown frontier in the physical sciences, in dealing with mysterious cosmological problems, we adopt a simple and effective approach, the Occam\rq s razor principle.

In line with studying the \textit{extended light field}, we use a unique hypothesis. The hypothesis is similar to the conformal cyclic cosmology (CCC) concept proposed by Penrose\cite{penro1}, whose characteristics were explained in Sakuma and Ojima\cite{sakuma3}, and hereinafter, it will be referred to as CCC of the second kind (CCC-2nd). According to it, as in the case of the creation and annihilation of the matter and antimatter pair, the twin universes as metric spacetimes are birthed by a certain kind of spontaneous conformal symmetry breaking of a light field with null distance ($ds^{2} = 0$). The twin universes thus created are divided by an event horizon intrinsically embedded in the de Sitter space structure, caused by the most dominant component of dark energy. Eons later, the twins meet at the event horizon to return to the original light field, and this cycle repeats forever. In our cosmological hypothesis, the flatness, isotropy, and horizon problems are resolved, respectively, by the observed existence ratio of dark energy to matter, the Weyl curvature hypothesis proposed by Penrose\cite{penro2}, and the existence of a superluminous off-shell electromagnetic field.

In this section, we first explain the emergence of a space-like domain known as de Sitter space mentioned above, which is necessary for quantum field interactions. The emergence mechanism of the familiar time-like domain to which not only the classical but also partial quantum fields belong must be quite different from that of the space-like domain. Second, we provide a clear explanation of the mechanism based on our new interpretation of conformal gravity. 

\subsection{Conformal symmetry breaking and emergence of spacetime}

Concerning the symmetry breaking of quantum fields, the MMD theory states that the symmetric space ($G/H$) studied by Cartan\cite{cart}, which has various desirable properties, emerges in the generalized sector classifying space if the symmetry of a given system, described by a Lie group ($G$), is broken up into its subgroup, $H$, remaining unbroken. A concrete example of this case will be explained in this subsection concerning the emergence of spacetime, \textit{not as a purely mathematical concept having nothing to do with physics, but as a physical entity.} Recall that $4D$ spacetime occupies a special position in differential topology. A most notable example would be the one highlighted by Donaldson \cite{dona}. In the context of our present research on QFT, the following properties of $4D$ spacetime are deemed important: \\ (i) the free Maxwell equation on electromagnetism is scale free only in $4D$ spacetime and (ii) the Weyl (conformal) curvature can be defined in spacetimes with dimensions larger than or equal to 4. 

If we combine the statement of the MMD theory on an emerging symmetric space ($G/H$) with property (i), i.e., the free Maxwell field that can play the role of $H$, then one of the assertions of our cosmological theory (CCC-2nd) proposing a $4D$ spacetime theory is restated as follows: \textit{the universe as a symmetric space ($G/H$) emerges as the conformal symmetry breaking of the $H$ field with $H$ being an important remaining unbroken subgroup}. In our series of papers cited in Section 1, we show how space- and time-like domains of $G/H$ emerge from $H$. Next, we briefly recapitulate the main points of those emergence processes, starting from the simple space-like case. 

\subsubsection{Emergence of the space-like domain}
The energy--momentum tensor ($T_{\mu}^{\;\;\nu}$) of the free Maxwell field assumes the form of $T_{\mu}^{\;\;\nu} = F_{\mu\sigma}F^{\nu\sigma}$ with $F_{\mu\sigma} = \nabla_{\mu}A_{\sigma}-\nabla_{\sigma}A_{\mu}$, where $\nabla_{\mu}$ denotes the covariant derivative and other notations are conventional. As is well known, this free field is characterized by a couple of conditions: 
\begin{equation}
F_{\nu\sigma}F^{\nu\sigma} = 0\;\;\;\; \textnormal{and}\;\;\;\; F_{01}F_{23} + F_{02}F_{31} + F_{03}F_{12} = 0. \label{eqn:5}
\end{equation}
In extending this null field to the space-like momentum domain, focusing on the properties of the Riemann curvature tensor ($R_{\alpha\beta\gamma\delta}$) and the mixed form of the Ricci tensor ($R_{\mu}^{\;\;\nu}$), as shown below, is quite helpful:
\begin{eqnarray}
R_{\beta\alpha\gamma\delta}& = &-R_{\alpha\beta\gamma\delta},\;\;R_{\alpha\beta\delta\gamma} = -R_{\alpha\beta\gamma\delta},\;\;R_{\gamma\delta\alpha\beta} = R_{\alpha\beta\gamma\delta},  \label{eqn:7} \\
R_{\alpha\beta\gamma\delta}& + &R_{\alpha\gamma\delta\beta} + R_{\alpha\delta\beta\gamma} = 0,\;\;
R_{\mu}^{\;\;\nu}: = R_{\mu\sigma}^{\;\;\;\;\nu\sigma}.  \label{eqn:9}
\end{eqnarray}
Here, the first equation in Eq. (\ref{eqn:9}) is called the first Bianchi identity, which is closely related to the second equation in Eq. (\ref{eqn:5}), namely, the orthogonality of electric and magnetic fields. 

In relativistic fluid dynamics, if the vorticity tensor field ($\omega_{\mu\nu}$) satisfies the second equation in Eq. (\ref{eqn:5}), then the fluid is classified as a barotropic (isentropic) fluid (cf. \cite{sakuma4}). This observation motivates us to use Clebsch parameterization (CP) to represent the electromagnetic vector potential ($U_{\mu}$) in space-like momentum domains. We can achieve this because the rotational part of the barotropic fluid velocity ($u_{\mu}$) can be represented by two Clebsch parameters, $\lambda$ and $\phi$, to give $u_{\mu} = \lambda \nabla_{\mu}\phi$ \cite{clebsch}. Notably, $\lambda$ and $\phi$ serve as the canonically conjugate variables for the Hamiltonian structure of the barotropic fluid motions. In our case, the two gradient vectors, $C_{\mu}: = \nabla_{\mu}\phi$ and $L_{\mu}: = \nabla_{\mu}\lambda$, with the orthogonality condition, $C^{\nu}L_{\nu} = 0$, provide a dynamical basis on which electric and magnetic fields can be successfully introduced into the space-like momentum domain.

To understand the meaning of CP, we first explain the light-like case to compare it with the conventional quantization of the light field. In the second part on the pure space-like case, we will show that the energy--momentum tensor ($\hat{T}_{\mu}^{\;\;\nu}$) in the extended space-like domain becomes isomorphic to the Einstein tensor in the theory of general relativity. This will reveal that the extended electromagnetic field automatically possesses the property of spacetime through which quantum electromagnetic field interactions occur (GR theorem). Namely, this spacetime, including the light-like case, plays the role of \lq\lq virtual photons'' in the conventional QED.  

\textbf{Light-like case}: $U^{\nu}(U_{\nu})^{*}=0$, where $(\cdot)^{*}$ denotes the complex conjugate of $(\cdot)$.\\
For Clebsch variables $\lambda$ and $\phi$ satisfying
\begin{eqnarray}
\nabla^{\nu}\nabla_{\nu}\lambda - (\kappa_{0})^{2}\lambda & = & 0,\;\;L_{\mu} = \nabla_{\mu}\lambda,\;\;\;\;\nabla^{\nu}\nabla_{\nu}\phi = 0,\;\;C_{\mu} = \nabla_{\mu}\phi,  \label{eqn:11}\\
C^{\nu}(L_{\nu})^{*} & = & 0,\;\;U_{\mu}: = \lambda C_{\mu},  \label{eqn:12}
\end{eqnarray}
where $(\kappa_{0})^{-1}$ is DP constant $l_{dp} \approx 50$ nm, determined by Ohtsu\cite{sakuma1}, we have
\begin{eqnarray}
U^{\nu}\nabla_{\nu}U_{\mu} = 0,\;\;S_{\mu\nu}: = \nabla_{\mu}U_{\nu}-\nabla_{\nu}U_{\mu} = L_{\mu}C_{\nu}
-L_{\nu}C_{\mu},  \label{eqn:13} \\
\hat{T}_{\mu}^{\;\;\nu} = S_{\mu\sigma}S^{\nu\sigma} = \rho C_{\mu}C^{\nu},\;\;\rho : = L^{\nu}(L_{\nu})^{*} < 0,\;\;
\nabla_{\nu}\hat{T}_{\mu}^{\;\;\nu} = 0. \label{eqn:15}  
\end{eqnarray}
Notice that $\hat{T}_{\mu}^{\;\;\nu}$ has a dual wave ($S_{\mu\sigma}S^{\nu\sigma}$) and particle ($\rho C_{\mu}C^{\nu}$) representation. The latter is consistent with the conventional quantization of QED where this mode is eliminated from the physically meaningful domain as the particle density ($\rho$) in Eq. (\ref{eqn:15}) becomes negative. As referenced in the introduction, in re-examining the Nakanishi--Lautrup formalism of the abelian gauge theory, Ojima\cite{ojima3} highlighted that the nonparticle form corresponding to ($S_{\mu\sigma}S^{\nu\sigma}$) in our model plays substantial physical roles, which is consistent with the GR theorem.

\textbf{Space-like case}: $U^{\nu}(U_{\nu})^{*} < 0$. \\
In this case, we modify Eq. (\ref{eqn:11}) as follows:
\begin{eqnarray}
\nabla^{\nu}\nabla_{\nu}\lambda - (\kappa_{0})^{2}\lambda & = & 0,\;\;\nabla^{\nu}\nabla_{\nu}\phi - (\kappa_{0})^{2}\phi = 0,\;\; C^{\nu}(L_{\nu})^{*} = 0,    \label{eqn:17} \\
U_{\mu}:& = & (\lambda C_{\mu} -\phi L_{\mu})/2.   \label{eqn:19}
\end{eqnarray}
By doing so, the associated energy--momentum tensor ($\hat{T}_{\mu}^{\;\;\nu}$) now becomes 
\begin{equation}
\hat{T}_{\mu}^{\;\;\nu} = S_{\mu\sigma}S^{\nu\sigma}-\frac{1}{2}g_{\mu}^{\;\;\nu} S_{\sigma\tau}S^{\sigma\tau}.
\label{eqn:21}
\end{equation}
Considering the definition of the Ricci tensor ($R_{\mu}^{\;\;\nu}$) in Eq. (\ref{eqn:9}), we see that Eq. (\ref{eqn:21}) is isomorphic to the Einstein tensor ($G_{\mu}^{\;\;\nu} = R_{\mu}^{\;\;\nu}-g_{\mu}^{\;\;\nu}R/2$).

As Dirac's equation ($(i\gamma^{\nu}\partial_{\nu} + m)\Psi = 0$) is the square root of the time-like KG equation ($(\partial^{\nu}\partial_{\nu} + m^{2})\Psi = 0$), the quantum field ($\Psi_{(M)}$) that satisfies the square root of the space-like KG equation ($(\partial^{\nu}\partial_{\nu}-(\kappa_{0})^{2})\Psi = 0$) is shown to be the electrically neutral Majorana fermionic field (\{$(\gamma^{\nu}_{(M)}\partial_{\nu}-\kappa_{0})\Psi_{(M)} = 0$\}).
As the space-like $\Psi_{(M)}$ represents a nonparticle mode, we denote it as the \textit{Majorana fermionic field} instead of the Majorana fermion. Owing to Pauli\rq s exclusion principle, $\Psi_{(M)}$ with a half-integer spin of 1/2 cannot occupy the same state. A possible configuration where a couple of $\Psi_{(M)}$s form a bosonic $S_{\mu\nu}$ field can be identified using the Pauli--Lubanski 4-vector ($W_{\mu}$), which describes the spin states of moving particles.
\begin{equation}
W_{\mu} = \frac{1}{2}\epsilon_{\mu\nu\lambda\sigma}M^{\nu\lambda}p^{\sigma},   \label{eqn:23}
\end{equation}  
where $\epsilon_{\mu\nu\lambda\sigma}$ denotes the $4D$ totally antisymmetric Levi--Civita tensor, $M^{\nu\lambda}$ and $p^{\sigma}$ are angular and linear momenta, respectively. As Eq. (\ref{eqn:23}) is rewritten as \\
\begin{eqnarray}
\left( \begin{array}{l}
    W_{0}  \\
    W_{1}  \\
    W_{2}  \\
    W_{3}
        \end{array}   \right)
=
\left( \begin{array}{cccc}
        0    &  M^{23}   &  M^{31}  &  M^{12}   \\
  -M^{23}  &    0       &  M^{03}  &  -M^{02}  \\
  -M^{31}  & -M^{03}  &     0      &  M^{01}  \\
  -M^{12}  &  M^{02}   & -M^{01} &    0
       \end{array}    \right)
\left( \begin{array}{r}
     p^{0}  \\
     p^{1}  \\
     p^{2}  \\
     p^{3}
       \end{array}   \right),  \label{eqn:25}
\end{eqnarray}
\\
\noindent we can see that two different fields, $(M^{\mu\nu}, p^{\mu}$) and ($N^{\mu\nu}, q^{\mu}$), which satisfy the orthogonality condition ($p^{\nu}q_{\nu} = 0$, corresponding to $C^{\nu}(L_{\nu})^{*} = 0$ in Eq. (\ref{eqn:17})), can share the same $W_{\mu}$ and hence combine to form a spin 1 bosonic field, $S_{\mu\nu}$.

The CDSE field ($S_{\mu\nu}$) has another noteworthy geometric feature, which is particularly important for its application in cosmology. A plane wave of the form, $\psi = \hat{\psi}_{c}\exp{[i(k_{\nu}x^{\nu})]}$, satisfies the equation of a couple of $\lambda$ and $\phi$ fields in Eq. (\ref{eqn:17}) with $\partial^{\nu}\psi(\partial_{\nu}\psi)^{*} = -(\kappa_{0})^{2}[\hat{\psi}_{c}(\hat{\psi}_{c})^{*}].$ Therefore, for vector $L_{\mu}$, we have
\begin{equation}
L^{\nu}(L_{\nu})^{*} = -(\kappa_{0})^{2}[\hat{\lambda}_{c}(\hat{\lambda}_{c})^{*}] = const. < 0,  \label{eqn:27}
\end{equation}
whose structure resembles that of the tangent vector on de Sitter space. Notably, the importance of this space in the context of spacetime quantization was first noted by Snyder\cite{hss}. A pseudo-hypersphere with radius $R_{dS}$ embedded in $R^{5}$ is called de Sitter space, which is represented as 
\begin{equation}
(\eta_{0})^{2}-(\eta_{1})^{2}-(\eta_{2})^{2}-(\eta_{3})^{2}-(\eta_{4})^{2} = -(R_{dS})^{2} = -3(\Lambda_{dS})^{-1}. \label{eqn:29}
\end{equation}
Here, $3(\Lambda_{dS})^{-1}: = (R_{dS})^{2}$ is a frequently employed alternative expression of $R_{dS}$.
Snyder showed that the spacetime ($x^{\mu}$) together with the 4-momentum ($p_{\mu}$), defined as 
\begin{equation}
p_{\mu} = \frac{\hbar}{l_{p}}\frac{\eta_{\mu}}{\eta_{4}},\;\;(0 \leq \mu \leq 3),\;\;
p^{\nu}p_{\nu} = \left(\frac{\hbar}{l_{p}}\right)^{2} \left[1-\left(\frac{R_{dS}}{\eta_{4}}\right)^{2}\right] < 0, 
\label{eqn:31}
\end{equation}
can be quantized without breaking Lorentz's invariance, where $l_{p}$ denotes Planck's length. We have already seen that an isomorphism exists between the Einstein tensor ($G_{\mu}^{\;\;\nu}$) and ($\hat{T}_{\mu}^{\;\;\nu}$) in Eq. (\ref{eqn:21}) whose r.h.s. can be quantized with the Majorana field. Thus, we conjecture that the spacetime quantization scheme of Snyder is consistent with the quantization of $\hat{T}_{\mu}^{\;\;\nu}$.
Eq. (\ref{eqn:27} and \ref{eqn:31}) show that $L_{\mu}$ and $p_{\mu}$ are on the submanifold of de Sitter space, parameterized by $\eta_{4} = const.$

As nonlinear electromagnetic field interactions are ubiquitous, the incessant occurrence of instantaneous excitation--deexcitation processes in the $S_{\mu\nu}$ field must be prevalent in the universe. In such processes, we can show that a unique state exists, $|M3\rangle_{g}$, of $S_{\mu\nu}$, which behaves as if it is \lq\lq the ground state\rq\rq of $S_{\mu\nu}$ in the sense that $|M3\rangle_{g}$ is occupied by $S_{\mu\nu}$ at every moment. To observe this phenomenon, we consider Eq. (\ref{eqn:23}) again. As the spatial dimension of our spacetime is three, the maximum number of space-like momentum vectors satisfying the orthogonality condition (\ref{eqn:17}): ($p^{\nu}q_{\nu} = 0$), is also three. That is, for three different sets of ($M^{\mu\nu},\; p^{\mu}$), ($N^{\mu\nu},\; q^{\mu}$), and ($L^{\mu\nu},\; r^{\mu}$) with $p^{\nu}q_{\nu} = 0;\;p^{\nu}r_{\nu} = 0;\;q^{\nu}r_{\nu} = 0$, the following unique spin vector configuration exists:
\begin{equation}
2W_{\mu} = \epsilon_{\mu\nu\lambda\sigma}M^{\nu\lambda}p^{\sigma} = \epsilon_{\mu\nu\lambda\sigma}N^{\nu\lambda}q^{\sigma} = \epsilon_{\mu\nu\lambda\sigma}L^{\nu\lambda}r^{\sigma}.   \label{eqn:33}
\end{equation}
This indicates the existence of a compound state of the Majorana fermionic field with spin 3/2, called the Rarita--Schwinger state ($|M3\rangle_{g}$). Notice that a bosonic field, $S_{\mu\nu}$, can be constructed by any pair in the set of \{($M^{\mu\nu},\; p^{\mu}$), ($N^{\mu\nu},\; q^{\mu}$), and ($L^{\mu\nu},\; r^{\mu}$)\} mentioned above. In our scenario of the electromagnetic field interactions in which the ubiquitous field interactions induce the incessant occurrence of excitation--deexcitation processes, the $|M3\rangle_{g}$ state behaves like \lq\lq the ground state\rq\rq of $S_{\mu\nu}$ in the sense that any vacant configuration of $S_{\mu\nu}$, if it exists, must be reoccupied in a moment. This makes $|M3\rangle_{g}$ a constantly occupied state from a macroscopic timescale.

The key question regarding $|M3\rangle_{g}$ is whether it is an observable quantity in either a direct or an indirect fashion as $S_{\mu\nu}$, as a space-like quantity, is not observable in general. First, we can say that, within the framework of relativistic QFT, any observable, without exception, associated with the given internal symmetry is the invariant under the action of the transformation group materializing the symmetry under consideration. The second helpful observation is that, as was highlighted after Eq. (\ref{eqn:31}), Lorentz's invariance still holds for Snyder\rq s quantum version of the isomorphism between Einstein tensor $G_{\mu}^{\;\;\nu}$ and the Majorana version of $\hat{T}_{\mu}^{\;\;\nu}$ given in Eq. (\ref{eqn:21}). This is in addition to the fact that Lorentz's invariance, indicating an external symmetry, is related to the internal ones through supersymmetry. Based on these observations, we assume that \textit{\{$|M3\rangle_{g}: = \Sigma_{(i=1)}^{(3)} \hat{T}_{\nu}^{\;\;\nu}(x^{i})\rightarrow R$\}, as the invariant of the general coordinate transformations, is an observable as it is indirectly related to the actual observable quantity, i.e., the expansion rate of the universe.}

The simplest model of dark energy is expressed as a cosmological term, $\Lambda_{(de)} g_{\mu\nu}$, with $\Lambda_{(de)} <0$ in the following Einstein field equation, with the sign convention $R >0$ for a matter-dominated closed universe:
\begin{equation}
R_{\mu}^{\;\;\nu}-\frac{R}{2}g_{\mu}^{\;\;\nu} + \Lambda_{(de)} g_{\mu}^{\;\;\nu} = -\frac{8\pi G}{c^{4}}T_{\mu}^{\;\;\nu}.   \label{eqn:35}
\end{equation}
From the discussions developed thus far, we can identify observable $\Lambda_{(de)}$ as
\begin{equation}
\Lambda_{(de)} = |M3\rangle_{g} = \Sigma_{(i=1)}^{(3)} \hat{T}_{\nu}^{\;\;\nu}(x^{i}) < 0.  \label{eqn:37}
\end{equation}
Recall that the key parameter ($\kappa_{0}$) in Eq. (\ref{eqn:11}) for determining Clebsch variables $\lambda$ and $\phi$ is the DP constant, which has been experimentally determined. Using the experimental value of $(\kappa_{0}^{-1})\approx 50$ nm, we obtain $\Lambda_{(de)} = -2.47\times 10^{-53} m^{-2}$. Conversely, the value of $\Lambda_{obs}$ derived by Planck satellite observations \cite{liu} is estimated as $\Lambda_{(obs)} \approx -3.7 \times 10^{-53} m^{-2}$. 

\subsubsection{Emergence of the time-like domain}
As shown above, the CP used to develop our dark energy model as a space-like electromagnetic field is an analytic approach that sheds light on the characteristics of \textit{barotropic (isentropic) fluid motions}. Next, we show that the knowledge on fluid dynamics is again quite helpful in identifying a promising dark matter model, which can be derived by considering \textit{baroclinic (nonisentropic) fluid motions.} For clarity, we begin with the definitions of barotropic and baroclinic fluids. For simplicity, let us consider the nonrelativistic equation of motion of ideal gas flows:
\begin{equation}
\partial_{t}v_{\mu} + v^{\nu}\partial_{\nu}v_{\mu} = -\partial_{\mu}p/\rho = -C_{p}\partial_{\mu}T + T\partial_{\mu}s,
\label{eqn:39}
\end{equation}
where $v_{\mu}$ denotes the $3D$ velocity field; $p$, $\rho$, $T$, and $s$ denote pressure, density, absolute temperature, and specific entropy, respectively; and $C_{p}$ denotes specific heat at constant pressure. A given fluid motion is barotropic if $\partial_{\mu}s = 0$ (namely isentropic); otherwise, it is baroclinic. A particularly important conserved quantity for the general baroclinic case is Ertel\rq s potential vorticity ($Q$\cite{ert}), which has the following form:
\begin{equation} 
Q: = \frac{1}{\rho}(\vec{\nabla}\times\vec{v})\cdot\vec{\nabla}s,\;\;\Longrightarrow\;\;\partial_{t}Q + v^{\nu}\partial_{\nu}Q = 0.  \label{eqn:41}
\end{equation}
Notice that $Q$ can be adopted to label the fluid particles defined in the Lagrangian specification of the flow field. Alternatively, $Q$ may play the role of the \lq\lq physical coordinates\rq\rq of the space in which a given baroclinic fluid system undergoes time evolution. 

To clarify the underlying mechanism of the interpretation of dark matter in terms of $Q$, Sakuma et al.\cite{sakuma4} examined the novel relativistic representation ($\Omega_{T}$) of $Q$. They found that \textit{under the assumption of a low energy limit within which the conservation of the fluid particle number ($n$) having the form of $\nabla_{\nu}(nu^{\nu}) = 0$ holds,} 
\begin{eqnarray} 
\Omega_{T}: = \Omega/T,\;\;\Omega: = \omega_{01}\omega_{23} + \omega_{02}\omega_{31} + \omega_{03}\omega_{12},   \label{eqn:43}
\\
\Omega_{T}u^{\mu} = \; \nabla_{\nu}[^{*}(\omega^{\mu\nu})(\sigma/n)],\;\;\Longrightarrow\;\;
\nabla_{\nu}(\Omega_{T}u^{\nu}) = 0.    \label{eqn:45}
\end{eqnarray}
Here, $u^{\mu}$, $\omega_{\mu\nu}$, $^{*}\omega^{\mu\nu}$, $T$, $\sigma$, and $n$ are the nondimensional 4-velocity vector, relativistic vorticity tensor, Hodge's dual of $\omega^{\mu\nu}$, absolute temperature, specific entropy, and particle number per unit volume, respectively. Furthermore, the current ($\Omega_{T}u^{\mu}$) is identified as gravitational entropy current. In additional, the conservation law given in the second equation in Eq. (\ref{eqn:45}) shows that the first equation may be considered as \lq\lq Maxwell's equation\rq\rq$\;$in which $\Omega_{T}u^{\mu}$ plays the role of the \lq\lq electric current\rq\rq. They further derived the following important relations valid among metric tensor $g^{\mu\nu}$, Weyl curvature tensor $W^{\alpha\beta\gamma\delta}$, and relativistic vorticity tensor $\omega^{\mu\nu}$ under the assumption of $W^{2}\Omega^{2}  \neq  0$
\begin{eqnarray}
g^{\mu\nu} = \frac{W^{\mu\alpha\beta\gamma}W^{\nu}_{\;\;\alpha\beta\gamma}}{W^{2}/4}
& = & \frac{^{*}{\omega}^{\mu\sigma}(^{*}{\omega}^{\kappa\lambda})\omega^{\nu}_{\;\;\sigma}\omega_{\kappa\lambda}}{(^{*}{\omega}^{\kappa\lambda}\omega_{\lambda\kappa})^{2}/4}
 = \frac{^{*}{\omega}^{\mu\sigma}(^{*}{\omega}^{\kappa\lambda})\omega^{\nu}_{\;\;\sigma}\omega_{\kappa\lambda}}{(4\Omega)^{2}/4},
\label{eqn:47} \\
W^{2}:& = &W^{\alpha\beta\gamma\delta}W_{\alpha\beta\gamma\delta},\;\;\Omega : = \;
^{*}{\omega}^{\kappa\lambda}\omega_{\lambda\kappa}. \nonumber
\end{eqnarray}
These suggest the existence of the minimum values of $(W_{0})^{2}$ and $(\Omega_{0})^{2}$.
In general, $g^{\mu\nu}$ is a purely mathematical concept as it is dependent on the choice of the coordinates. \textit{However, Eq. (\ref{eqn:47}) assigns a unique physical characteristic to $g^{\mu\nu}$ such that the emergent symmetric space ($G/H$) in our cosmological theory, as a metric space, assumes the form of a certain kind of spin-network to which the skew-symmetric properties of Eq. (\ref{eqn:7}) are reflected.} In relativistic dynamics, mass distribution is directly related to the distribution of scalar curvature $R_{\nu}^{\;\;\nu}$. By contrast, from Eq. (\ref{eqn:47}), we observe that $W^{2}$ is directly correlated with $(4\Omega)^{2}$ and that $\Omega_{T}u^{\mu} = (\Omega/T)u^{\mu}$ behaves like a conserved density current. As the pure Weyl curvature represents the vacuum of relativistic dynamics, a relatively strong current ($\Omega_{T}u^{\mu}$) in a cosmological environment with a negligible value of $R_{\nu}^{\;\;\nu}$ compared with $\Omega_{T}$ would behave like the invisible density current, which is the model of dark matter in our theory. Using Aoki et al.\rq s\cite{aoki2} recent research outcomes on relativistic conserved charges and entropy current, Sakuma et al.\cite{sakuma4} also showed that \textit{the entropy current $\Omega_{T}u^{\mu}$ is dynamically associated with an energy--momentum tensor of the form, $\lambda g_{\mu\nu}$, with $\lambda$ being a positive constant, where $g_{\mu\nu}$ is to be interpreted by Eq. (\ref{eqn:47})}.   

Thus, a series of the above analyses on $\Omega_{T}u^{\mu}$ suggests that the energy--momentum tensor of dark matter assumes the form of $\Lambda_{(dm)} g_{\mu\nu}$, where constant $\Lambda_{(dm)}$ is to be determined by observations. The consensus ranges of the estimated percentage of dark energy and dark matter are (68\%--76\%; mean = 72\%) and (20\%--28\%; mean = 24\%), respectively. Thus, the following equation:
\begin{equation}
\Lambda_{(dm)} \approx \hat{\Lambda}_{(de)}: = -\Lambda_{(de)}/3 = \Lambda_{dS}/3 = 1/(R_{dS})^{2},  
\label{eqn:49}
\end{equation} 
is an estimate consistent with the observation, where $R_{dS}$ and $\Lambda_{dS}/3$ denote the radius of de Sitter space and its alternative expression defined in Eq. (\ref{eqn:29}). 

Notably, in our cosmological scenario of CCC-2nd, $\hat{T}_{\mu}^{\;\;\nu}$ in Eq. (\ref{eqn:21}), shown to be isomorphic to the space-like Einstein tensor ($G_{\mu}^{\;\;\nu}$) and $\Lambda_{(dm)} g^{\mu\nu}$ with $g^{\mu\nu}$ given by Eq. (\ref{eqn:47}), provide \textit{the physical} space- and time-like spacetimes, respectively. These spacetimes emerged from the spontaneous symmetry breaking of the \textit{primordial} light field with null distance ($ds^{2} = 0$). To consolidate our CCC-2nd scenario here, we show that a unique vector boson exists in our model, which plays the role of the Nambu--Goldstone boson (NGB) associated with the spontaneous symmetry-breaking process. 

To start, we consider again the space-like CP of $U^{\mu}$ explained in Eq. (\ref{eqn:17}, \ref{eqn:19}).
The field strength ($S_{\mu\nu}$) defined by the curl of $U_{\mu}$ assumes the same form as the one in the light-like case given in the second equation in Eq. (\ref{eqn:13}). Contrarily, the null geodesic equation ($U^{\nu}\nabla_{\nu}U_{\mu} = 0$) in Eq. (\ref{eqn:13}) is replaced by
\begin{equation}
U^{\nu}\nabla_{\nu}U_{\mu} = -S_{\mu\nu}U^{\nu} + \nabla_{\nu}(U^{\nu}U_{\nu}/2) = 0;\;\; U^{\nu}U_{\nu} < 0.  \label{eqn:307}
\end{equation}
In relativistic fluid dynamics, the magnitude of $U^{\mu}$, defined by $V: = U^{\nu}U_{\nu}/2$, can be normalized as $V = 1$\cite{Landau}. As the equations on $\lambda$ and $\phi$ in Eq. (\ref{eqn:17}) are linear, we can introduce a similar normalization for $L_{\mu} = \nabla_{\mu}\lambda$ and $C_{\mu} = \nabla_{\mu}\phi$. The natural normalization would be
\begin{equation}
L^{\nu}L_{\nu} = -(m_{\lambda})^{2}\lambda^{2},\;\;C^{\nu}C_{\nu} = -(m_{\phi})^{2}\phi^{2}, \label{eqn:301}
\end{equation}
where the wavenumber vector, $k^{\mu}$, of the respective plane-wave solutions satisfies $k^{\nu}k_{\nu} = -(m_{\lambda})^{2}$ and $k^{\nu}k_{\nu} = -(m_{\phi})^{2}$. For the space-like CP of $U^{\mu}$ discussed thus far, we have 
\begin{equation}
(m_{\lambda})^{2} = (m_{\phi})^{2} = (\kappa_{0})^{2}.   \label{eqn:303}
\end{equation}

Using the orthogonality condition, $C^{\nu}(L_{\nu})^{*} = 0$, in Eq. (\ref{eqn:17}) and Eq. (\ref{eqn:301}), a couple of important characteristics of space-like $U^{\mu}$ fields can readily be expressed as
\begin{equation}
\nabla_{\nu}U^{\nu} = 0,\;\;\;V = -\frac{1}{8}\lambda^{2}\phi^{2}[(m_{\lambda})^{2} + (m_{\phi})^{2}].  \label{eqn:305} 
\end{equation}

Now, revisiting Eq. (\ref{eqn:17}), we consider a different case in which we only replace the second equation, $\nabla^{\nu}\nabla_{\nu}\phi - (\kappa_{0})^{2}\phi = 0$, with
$\nabla^{\nu}\nabla_{\nu}\phi + (\kappa_{0})^{2}\phi = 0$. With this change, Eq. (\ref{eqn:303})
changes into $(m_{\lambda})^{2} = (\kappa_{0})^{2} = -(m_{\phi})^{2}$. Consequently, $V$ in Eq. (\ref{eqn:305}) vanishes, and $U^{\mu}$ becomes a null vector. Notably, the form of $S_{\mu\nu}U^{\nu}$ in Eq. (\ref{eqn:307}) is similar to that of Lorentz force, $F_{\mu\nu}(ev^{\nu})$, where $F_{\mu\nu}$ and $ev^{\nu}$ denote the background electromagnetic field and an electric current with charge $e$, respectively. A direct expression of $-S_{\mu\nu}U^{\nu}$ is
\begin{eqnarray}
-S_{\mu\nu}U^{\nu} & = & -(L_{\mu}C_{\nu}-L_{\nu}C_{\mu})(\lambda C^{\nu} - \phi L^{\nu})/2  \nonumber \\ 
& = & -\lambda \phi (\kappa_{0})^{2}(\phi L_{\mu} - \lambda C_{\mu})/2 = \lambda\phi (\kappa_{0})^{2}U_{\mu}. \label{eqn:309}
\end{eqnarray}
Conversely, as we have $\nabla_{\nu}U^{\nu} = -\lambda\phi (\kappa_{0})^{2}$, in this case, Eq. (\ref{eqn:309}) becomes
\begin{equation}
S_{\mu\nu}U^{\nu} = -\lambda\phi (\kappa_{0})^{2}U_{\mu} = (\nabla_{\nu}U^{\nu})U_{\mu} \neq 0. \label{eqn:311}
\end{equation}
Eq. (\ref{eqn:311}) expresses that \textit{the null vector $U^{\mu}$ with a nonvanishing irrotational part} is a vector field for which $\kappa_{0}$ and $i\kappa_{0}$ play the role of the field source as in the case of $\pm e$ in electromagnetism. Furthermore, the field source behaves as if it is an \lq\lq electrically charged virtual photon'' responsible for nonlinear electromagnetic field interactions. In our CCC-2nd scenario, the emergence of a couple of these sources ($\kappa_{0}$ and $i\kappa_{0}$) is interpreted as the consequence of the spontaneous symmetry breaking of the light vector field whose temporal and spatial components exist in a balanced manner. Thus, we believe that the vector field, $U^{\mu}$, with the abovementioned characteristics plays the role of NGB in our cosmological scenario. Moreover, it plays substantial roles in the nonlinear field interactions between the quantum generalized sectors defined by Ojima\cite{ojima1}, existing on not only time-like but also space-like domains, and the classical generalized sectors on time-like domains. In the subsequent section, we show that the DP constant ($l_{dp} \approx 50$ nm), defined as the inverse of $\kappa_{0}$, gives the scale of the Heisenberg cut dividing the whole universe into micro-quantum and macro-classical worlds (cf. Eq. (\ref{eqn:207})).

Interestingly, we can unravel many conceptual aspects of QFT from such a specific theory as the description of the behaviors of DPs. Here, the concentration of the momentum support of the dressed photon field in the space-like domain ($p^{2} = p_{\mu }^{\;\;}p^{\mu } < 0$) plays the most important role, which is in sharp contrast to the time-like momenta owing to the particle-like excitations. This contrast between the space- and time-like momenta should not be superficially considered. However, the important factor is the medium or environment constructed by the space-like momenta, in contrast to the particle-like motions associated with the time-like momenta. From this viewpoint, the distribution of different sizes of space-like momenta ($p^{2} < 0$) describes the difference among different media, which constitutes the categorical background of algebraic QFT constructed by arrows of space-like momenta. Therefore, this kind of interpretation of space-like momenta associated with dressed photons exhibits the categorical essence of QFT.

Along this line, we promote new research on the relevance of Kan extentions in the QFT of dressed photons. This is applicable if the four terms constituting MMD are recombined into the three terms of micro-dynamical systems (comprising micro-dynamics acting on the micro-algebra) and macro-states, together with a macro-classifying space constructed by the classifying parameters of macro-states.

\section{Approach to the hierarchy problem}
The Einstein field equation incorporating our dark energy and matter model is expressed as follows:
\begin{eqnarray}
R_{\mu}^{\;\;\nu}-\frac{R}{2}g_{\mu}^{\;\;\nu} + \Lambda_{(dm)} g_{\mu}^{\;\;\nu} & = & -\frac{8\pi G}{c^{4}}[T_{\mu}^{\;\;\nu} + \{\Lambda_{(de)}g_{\mu}^{\;\;\nu}\}],   \label{eqn:191}  \\
\Lambda_{(dm)} & = & -\Lambda_{(de)}/3 + \epsilon > 0. \nonumber
\end{eqnarray}
Here, our present concern is the dark matter field, $\Lambda_{(dm)} g_{\mu}^{\;\;\nu}$, as the main source of the gravitational field. As the first step, by comparing Coulomb\rq s law with the universal law of gravitation, we obtain
\begin{equation}
F_{e} = \frac{1}{4\pi \epsilon_{0}}\frac{q_{1}q_{2}}{r^{2}},\;\;\;F_{g} = G\frac{m_{1}m_{2}}{r^{2}}.  \label{eqn:193}
\end{equation} 
Let us examine the $F_{e}/F_{g}$ ratio, where the notations used in Eq. (\ref{eqn:193}) are conventional. By doing so, it would be natural to choose the fundamental electric charge ($e$) for $q_{1} = q_{2}$. However, for $m_{1} = m_{2}$, we encounter a serious challenge as we cannot single out \lq\lq the fundamental mass charge\rq\rq$\;$like $e$ in the case of $F_{e}$. To overcome this problem, recall first that, for our dark energy model, a unique state, $|M3\rangle_{g}$, exists, which behaves like \lq\lq the ground state\rq\rq$\;$of $S_{\mu\nu}$. Through Eq. (\ref{eqn:37}) and Eq. (\ref{eqn:49}), this \lq\lq ground state\rq\rq$\;$is directly related to $\Lambda_{(dm)}$. By contrast, in Eq. (\ref{eqn:47}), we highlight the existence of the minimum value of $W^{2}$, i.e., $(W_{0})^{2} \neq 0$, associated with the conformal symmetry breaking of the light field ($H$). Thus, we naturally assume that $\Lambda_{(dm)} = |W_{0}|$. Notably, in our previous studies where we did not consider the possibility of the \lq\lq temporal change\rq\rq$\;$of $\Lambda_{(dm)}$ in the dark matter field ($\Lambda_{(dm)}g_{\mu\nu}$), we simply asserted $\Lambda_{(dm)} = W_{0} = const$. However, in the present discussion where we consider the \lq\lq temporal change\rq\rq$\;$of $\Lambda_{(dm)}$, we assume that $W_{0}$ is a certain positive constant that satisfies
\begin{equation}
W_{0}: = Min\{\Lambda_{(dm)}\}.  
\end{equation}
Under this new hypothesis, we can regard $\Lambda_{(dm)}$ as the fundamental mass of the gravitational field; the justification will be provided shortly.

As the dimension of $\Lambda_{(dm)}$ is $(\textnormal{length})^{-2}$, we introduce $m_{\lambda}$ having the dimension of mass, corresponding to $\Lambda_{(dm)}$. Substituting $e$ and $m_{\lambda}$ into $F_{e}$ and $F_{g}$ in Eq. (\ref{eqn:193}), respectively, we obtain
\begin{eqnarray}
\frac{F_{e}}{F_{g}} & = & \frac{e^{2}}{4\pi\epsilon_{0}}\frac{1}{G(m_{\lambda})^{2}} = \frac{e^{2}}{4\pi\epsilon_{0}c\hbar}\frac{c\hbar}{G(m_{\lambda})^{2}} = \alpha\left(\frac{m_{p}}{m_{\lambda}}\right)^{2},  \label{eqn:195}  \\
\alpha &: = & \frac{e^{2}}{4\pi\epsilon_{0}c\hbar},\;\;\;m_{p}: = \sqrt{\frac{c\hbar}{G}}, \label{eqn:197} 
\end{eqnarray}
where $\alpha$ and $m_{p}$ are the coupling constant of the electromagnetic field and Planck's mass.
For $\Lambda_{(dm)}$, using Eq. (\ref{eqn:49}), namely, $\Lambda_{(dm)}\approx -\Lambda_{(de)/3}$, and 
using the concrete expression of Eq. (\ref{eqn:37}) obtained in reference [17], we have
\begin{equation}
\Lambda_{(dm)} \approx \frac{4\pi Gh}{c^{3}}\frac{(\kappa_{0})^{2}}{\epsilon} = \frac{8\pi^{2}G\hbar}{c^{3}}\frac{(\kappa_{0})^{2}}{\epsilon} = 8\pi^{2}(l_{p})^{2}\frac{1}{\epsilon (l_{dp})^{2}};\;\;l_{dp}: = (\kappa_{0})^{-1}. 
\label{eqn:199} 
\end{equation}
Here, $l_{p}$ and $l_{dp}$ are the Planck length and DP length defined as the inverse of the \textit{DP constant} in Eq. (\ref{eqn:11}); $\epsilon$ denotes a dimension adjusting coefficient of \textit{unit length squared}. Two reasons exist for the appearance of $\epsilon$ in the expression of $\Lambda_{(dm)}$. First, the quantity, $\Lambda_{(de)} = \Sigma_{(i=1)}^{(3)} \hat{T}_{\nu}^{\;\;\nu}(x^{i})$, in Eq. (\ref{eqn:37}) is related to the \lq\lq radiation pressure\rq\rq$\;$of the $S_{\mu\nu}$ field. Second, the calculation of such a quantity is required to make the energy quantization of the light-like $S_{\mu\nu}$ field consistent with $E = h\nu$ for the usual radiation field. As we have introduced $m_{\lambda}$ as the elemental mass corresponding to $\Lambda_{(dm)}$, we can determine it by the following formal identification using Einstein's field equation: 
\begin{equation}
\Lambda_{(dm)}g_{\mu}^{\;\;\nu} = \frac{8\pi G}{c^{4}}[(\rho_{\lambda}c^{2})u_{\mu}u^{\nu}],\;\;\Rightarrow\;\;\Lambda_{(dm)}(l_{\epsilon})^{3} = \frac{8\pi G}{c^{2}}m_{\lambda},\;\;m_{\lambda} = \rho_{\lambda}(l_{\epsilon})^{3},
\label{eqn:201}
\end{equation}
where $l_{\epsilon}$ denotes unit length. Therefore, using Eq. (\ref{eqn:195}), Eq. (\ref{eqn:197}), Eq. (\ref{eqn:199}), and Eq. (\ref{eqn:201}), we finally obtain
\begin{equation}
\frac{F_{e}}{F_{g}} = \frac{\alpha}{\pi^{2}}\frac{(l_{dp})^{4}}{(l_{p})^{2}}\frac{1}{\epsilon}.   \label{eqn:203}
\end{equation}
Substituting $\alpha = 7.3\times 10^{-3}$, $\pi^{2} = 9.9$, $l_{p} = 1.6\times 10^{-35}$m, $l_{dp} \approx 5.0\times 10^{-8}$m, and $\epsilon = 1$$\textnormal{m}^{2}$ into Eq. (\ref{eqn:203}), we obtain
\begin{equation}
\frac{F_{e}}{F_{g}} = 1.7 \times 10^{37},  \label{eqn:205}
\end{equation}
which appears to be consistent with the conventional rough estimates obtained without using $\Lambda_{(dm)}$. 

Having derived Eq. (\ref{eqn:203}), now we examine the consequence of the \lq\lq temporal change'' of $R_{dS}(= 1/\sqrt{\Lambda_{(dm)}})$ applicable to $F_{e}/F_{g}$ in Eq. (\ref{eqn:203}). In the introduction, we cited previous studies on (cosmological) time-dependent physical constants. Suppose that $\Lambda_{(dm)}$ is such a time-dependent quantity, then $\Lambda_{(dm)}g_{\mu}^{\;\;\nu}$ in Eq. (\ref{eqn:191}) ceases to be a divergence-free term. Notice, however, that as $\Lambda_{(dm)}$ is directly related to the radius ($R_{dS}$, in Eq. (\ref{eqn:49})) of de Sitter space (Eq. (\ref{eqn:29})), the partial derivative of $\Lambda_{(dm)}$ with respect to the cosmological coordinate ($x^{\mu}$) vanishes. This is because the shape of the isotropic universe is given by a de Sitter space and $\nabla_{\mu}R_{dS} \; (0\leq \mu \leq 3)$ is on the tangent hyperplane of $R_{dS} = const.$, on which the gradient of the local $4D$ spacetime coordinate ($x^{\mu}$) exists. Thus, in this sense, $\Lambda_{(dm)}g_{\mu}^{\;\;\nu}$ remains a divergence-free term, although the radius can either expand or shrink in the fifth dimension perpendicular to the gradient of the $4D$ coordinates ($x^{\mu}$). The \lq\lq temporal change'' of $\Lambda_{(dm)}$ that we consider is the change in the magnitudes of the dark energy and matter fields. 

For simplicity, in our discussion, except for $\Lambda_{(de)}$ and the related $\Lambda_{(dm)}$, we assume that all physical constants are fixed quantities. From Eq. (\ref{eqn:49}, \ref{eqn:199}), we readily obtain
\begin{equation}
l_{dp} \approx \frac{2\sqrt{2}\pi}{\sqrt{\epsilon}}\frac{l_{p}}{\sqrt{\Lambda_{(dm)}}}
 = \frac{2\sqrt{2}\pi}{\sqrt{\epsilon}}l_{p}R_{dS},  \label{eqn:207}
\end{equation}
which implies that the DP length ($l_{dp}$) affords the geometric mean of the smallest Planck length ($l_{p}$) and the largest scale of our universe ($R_{dS}$). Furthermore, under the assumption that $l_{p} = const.$, $l_{dp}$ becomes proportional to $R_{dS}$. Thus, we can choose $l_{dp}$ as the unique geometrical parameter of our cosmological model. From the viewpoint of the unification of four forces, examining the case where we have $F_{e}/F_{g} = 1$ is interesting. A simple calculation shows that
\begin{eqnarray}
\textnormal{the present value}:\;\;\frac{F_{e}}{F_{g}} & = & 1.7\times 10^{37};\;\;l_{dp} 
\approx 5.0\times 10^{-8}\textnormal{m}  \label{eqn:209}  \\   
\textnormal{the unification value}:\;\;\frac{F_{e}}{F_{g}} & = & 1;\;\;\;\;\;\;\;\;\;\;\;\;\;\;\;\;l_{dp} \approx 2.4 \times 10^{-17}\textnormal{m}.
\end{eqnarray}  
Thus, using Eq. (\ref{eqn:199}), we observe that the unification value of $\Lambda_{(dm)}(u)$ is ($4.4\times 10^{18}$) times larger than the present value of $\Lambda_{(dm)}(p)$.

We focus on dark energy and matter mainly because of their extreme predominance over material substances. This implies that, in thermodynamic terminology, the dark matter field ($\Lambda_{(dm)}g_{\mu\nu}$) with positive $\Lambda_{(dm)}$ resulting from \textit{Weyl curvature} and dark energy field ($\Lambda_{(de)}g_{\mu\nu}$) with negative $\Lambda_{(de)}g_{\mu\nu}$ resulting from \textit{Ricci curvature} would work as high- and low-temperature reservoirs, respectively, for the gravity-driven temporal evolution of such material systems as stars, galaxies, clusters of galaxies, and the large-scale structure of the cosmos. Moreover, in this cosmic thermodynamical system, $\Omega_{T}u^{\mu}$, defined in Eq. (\ref{eqn:45}), gives the gravitational entropy flow. As the energy density of the two thermal reservoirs are finite, the initial \lq\lq temperature difference'' between them, measured by $\Lambda_{(dm)}(u)-\Lambda_{(de)}(u) \approx 4\Lambda_{(dm)}(u)/3 = 4[R_{dS}(u)]^{-2}/3\;$, would decrease with the temporal evolution of material systems. This result can be observed as the extra expansion (increase of $R_{dS}$) of our universe in the fifth dimension, directly related to the temporal increase in the ratio of the previously discussed coupling constants, $F_{e}/F_{g} \propto (R_{dS})^{2}$. Although the dynamics we have discussed are unrelated to superstring theories, it is interesting to bring our attention to Witten\rq s noteworthy remark\cite{witten} made at \textit{Strings \lq 95}: \lq\lq eleven-dimensional supergravity arises as a low energy limit of the ten-dimensional Type IIA superstring.'' This appears to be qualitatively similar to our present situation, in which our $4D$ universe undergoes an extra expansion into the surrounding fifth-dimensional space. The expansion starts from the initial high-energy state of $F_{e}/F_{g} \approx 1$ with a negligible magnitude of $W^{2}$ in Eq. (\ref{eqn:47}) to low energy states having large values of $W^{2}$ as the measure of conformal gravity.

\section{Implication of Maldacena (AdS/CFT) duality}
In subsection 2.1.1, we saw that the CDSE field ($S_{\mu\nu}$) is closely related to de Sitter space having the well-known scale-free property. Recall first that spinor is an irreducible representation of the universal covering group, $SU(2)$ of $SO(3)$. For the $4D$ spacetime case, in which we have the Lorentz transformation group ($SO(1,3)$), the Lorentzian spinor ($SL(2,C)$) corresponds to $SU(2)$ in the case of $SO(3)$. When we further extend $SO(1,3)$ into the $4D$ conformal transformation group ($SO(2,4)$), $SL(2,C)$ is extended into $SU(2,2)$, which operates on Penrose\rq s twistor in the $4D$ complex spacetime. As in the case of the above extension of spinor, we can also consider a similar extension of the electromagnetic field ($F_{\mu\nu}$) as the $U(1)$ gauge field. We believe that the CP (explained in subsection 2.1.1) applied to extend $F_{\mu\nu}$ into a space-like momentum domain is what is required for such an extension of conformal transformation, which is closely related to the important notion of modular form.  

The fact that the emergence of de Sitter space through CP is an inevitable consequence of the extension of
$SL(2,C)$ can be readily verified from the following properties of the Lorentzian spinor, $\Psi (V^{\mu})$:
\begin{eqnarray}
\qquad \Psi (V^{\mu}) = V^{\mu(\mu)^{\prime}}& = &\left[ 
\begin{array}{cc}
V^{00^{\prime}} & V^{01^{\prime}} \\ 
V^{10^{\prime}} & V^{11^{\prime}}
\end{array}
\right] = \frac{1}{\sqrt{2}}\left[ 
\begin{array}{cc}
V^{0} + V^{3} & V^{1} + iV^{2} \\ 
V^{1}-iV^{2} & V^{0}-V^{3}
\end{array}
\right],  \label{eqn:51} \\
det \Psi (V^{\mu}) & = & \frac{1}{2}[(V^{0})^{2}-(V^{1})^{2}-(V^{2})^{2}-(V^{3})^{2}].  \label{eqn:53}
\end{eqnarray}
For the space-like vector, $V^{\mu}$, $det \Psi (V^{\mu})$ becomes negative so that Eq. (\ref{eqn:53}) becomes isomorphic to the second equation in Eq. (\ref{eqn:31}). As Eq. (\ref{eqn:29}) and Eq. (\ref{eqn:31}) are connected by a one-to-one correspondence through parameter $\eta_{4}$ and the latter is isomorphic to Eq. (\ref{eqn:53}), we thus see that de Sitter space and $\Psi (V^{\mu})$ share the same symmetry. 

Now, we compare the following forms of $g^{\mu\nu}$:
\begin{eqnarray}
g^{\mu\nu}_{(I)} & = & \frac{-1}{\Lambda_{(I)}}\left(R_{(I)}^{\mu\nu}-\frac{R_{(I)}}{2}g^{\mu\nu}_{(I)}\right),\;\;(W^{\alpha\beta\gamma\delta} = 0),\;(I=1\;\textnormal{or}\; 2), \label{eqn:55}\\
g^{\mu\nu} & = & \frac{W^{\mu\alpha\beta\gamma}W^{\nu}_{\;\;\alpha\beta\gamma}}{W^{2}/4}\;\; (W^{2} \neq 0),
\label{eqn:57}
\end{eqnarray}
where Eq. (\ref{eqn:55}) represents either de Sitter space with $\Lambda_{(1)} = \Lambda_{(de)} < 0$ or anti de Sitter space (AdS) with $\Lambda_{(2)} = \Lambda_{(dm)} > 0$, depending on the index ($I$). 

In our CCC-2nd hypothesis, the isotropy of an early universe is explained by the small amplitude of $W^{2}$ (the Weyl curvature hypothesis proposed by Penrose\cite{penro2}), which motivates us to examine the possibility that the following limit exists:
\begin{eqnarray}
g^{\mu\nu} = \frac{W^{\mu\alpha\beta\gamma}W^{\nu}_{\;\;\alpha\beta\gamma}}{W^{2}/4}\;\longrightarrow \; g^{\mu\nu}_{(2)} = \frac{-1}{\Lambda_{(2)}}\left(R_{(2)}^{\mu\nu}-\frac{R_{(2)}}{2}g^{\mu\nu}_{(2)}\right),
\;\textnormal{as}\;\;W^{2}\;\rightarrow\;0.  \label{eqn:59}
\end{eqnarray} 
Within the framework of our CCC-2nd hypothesis, this gives a diffeomorphism defined as the time reversal of a given cosmological time development. As $g^{\mu\nu}$ on the l.h.s. gives the gravitational field, whereas the r.h.s. represents AdS, Eq. (\ref{eqn:59}) can be regarded as the \lq\lq Maldacena (AdS/CFT) duality\rq\rq \cite{mald} in CCC-2nd. 

\section{Summary and conclusion}
Using our recently developed unconventional model incorporating dark energy and matter, we re-examined the hierarchy problem. As the model itself is not well known, we began our discussion by recapitulating the key concepts used in our new theory to elucidate an elusive phenomenon called \textit{DP} in the field of nanophotonics. This phenomenon unexpectedly inspired the novel perspective on cosmological problems addressed in this study. The concepts include the GR theorem for nonlinear quantum field interactions, Ojima\rq s MMD theory developed from DHR\rq s original sector theory in QFT, and CDSE field developed by the first author, together with the unconventional introduction of conformal gravity explained in subsection 2.1.

In addition, as an outstanding feature of our cosmological theory, we use the basic view that our universe is undergoing an infinite cycle of birth and death as in the case of CCC. However, a crucial difference exists between CCC and our theory (CCC-2nd), namely, the twin structure of a universe arising from the intrinsic property of de Sitter space. Our reason for supporting such a view is twofold. First, we believe that the basic principle of the universe is not complicated but simple. The creation and annihilation mechanisms of the matter and antimatter pair through the intervention of a light field observed in laboratories can also be applied naturally to the case of a twin universe configuration. Once we accept this conceptually simple view on the creation of the universe, we need not worry about complicated parameter tuning processes used in the widely prevailing theory of cosmic inflation, which favors the creation of everything from emptiness. 

Second, as mentioned in the introductory section, the idea of the creation of everything from emptiness appears to be conceived from either the misinterpretation or the extended interpretation of Fock vacuum $|0\rangle$, which is, as far as QFT is concerned, introduced within the framework of linear dynamics on free modes. In our cosmological theory, as Eq. (\ref{eqn:49}) shows, the universe expands, keeping a quasi-equilibrium between $\Lambda_{(dm)}$ and $\hat{\Lambda}_{(de)}$, which behave as \lq\lq high and low temperature reservoirs'', respectively, for the temporal evolutions of systems in the universe. Thus, the essential global aspect of the expanding universe in a quasi-equilibrium state should be described by the Tomita--Takesaki theory\cite{tom} as a thermodynamic Kubo--Martin--Schwinger (KMS) state with infinite degrees of freedom. As the KMS state is a mixed one, its corresponding Gel\rq fand--Naimark--Segal representation is reducible. Therefore, for $\mathcal{M}$, defined as a von Neumann algebra on Hilbert space $\mathfrak{H}$, there exists its commutant, $\mathcal{M}^{\prime}$, which satisfies the following inversion relation:
\begin{eqnarray}
J \mathcal{M}J = \mathcal{M}^{\prime},\;\;\; e^{itH}\mathcal{M} e^{-itH} = \mathcal{M},\;\;\;J^{2} = 1,  \label{eqn:211} \\
J H J = -H,    \label{eqn:213}
\end{eqnarray}
where $H$ and $J$ denote the Hamiltonian and antiunitary operators called modular conjugation, respectively. Notice that the spectrum of the Hamiltonian is symmetric with respect to its sign, which indicates the existence of states with negative energy. We believe that the result of the Tomita--Takesaki theory applies to the case of a twin universe configuration and to the thermodynamics of the observed cosmic background radiation whose energy-spectrum distribution is given by the black body radiation. 

Eq. (\ref{eqn:203}) stands as the central result in our study, showing the clear causal relation between the existence of noncompactified extra dimension and $F_{e}/F_{g}$, in which the DP length ($l_{dp}$) plays the key role in determining the value of $F_{e}/F_{g}$. Through Eq. (\ref{eqn:207}), we see that $l_{dp}$ divides the universe into two domains. Furthermore, its present value, given in Eq. (\ref{eqn:209}), suggests that $l_{dp}$ affords the scale of the Heisenberg cut dividing the micro-quantum and the macro-classical worlds. 

As the final remark on our study, we wish to make a few brief comments on the innovative observational outcomes of the James Webb Space Telescope (JWST). One of the quite unexpected findings of the \textit{JWST Advanced Deep Extragalactic Survey} is that the events considered to have occurred in the very early stages of the universe actually occurred much earlier than expected. We have shown that the magnitudes of $\Lambda_{(dm)}$ and $\Lambda_{(de)}$ given in Eq. (\ref{eqn:49}) in the early stage of the universe were considerably greater than those in the present one. Thus, we conjecture that the standard $\Lambda$CDM model used in cosmological simulations is not suitable for simulating the temporal development of the early universe. Eq. (\ref{eqn:191}) shows that the isotropic expansion owing to $\Lambda_{(de)}g_{\mu\nu}$ on the r.h.s. manifests itself through a conformally flat Ricci curvature. Conversely, the local gravitational attraction owing to $\Lambda_{(dm)}g_{\mu\nu}$ on the l.h.s. is highly nonisotropic because of the intrinsic property of the Weyl curvature. Data from CEERS survey by JWST show that most of the very early galaxies assumed elongated shapes similar to noodles or surfboards, which appears to be consistent with our conjecture on the gravitational effect arising from the nonisotropic Weyl curvature. Presumably, the observed large-scale structure of the present universe comprising galaxy filaments and voids is also the manifestation of this nonisotropic Weyl curvature effect.

Regarding the cosmic age problem relating to the very early formation of primordial galaxies, in his 2023 paper, Gupta\cite{gupta} proposed an intriguing resolution derived from the outcome of a hybrid cosmological model called CCC(\textit{covarying coupling constants}) + TL(\textit{tired light}). The idea of CCC is similar to that of Dirac\rq s hypothesis on varying physical constants, such as gravitational constant ($G$) and speed of light ($c$). Gupta introduced the CCC model as an extended version of the $\Lambda$CDM model with \textit{a variant cosmological constant}. In that sense, our model with a varying $\Lambda_{(dm)}(\approx -\Lambda_{(de)}/3)$ is similar to the CCC model. However, a crucial difference exists between these two models. In Gupta\rq s model, parameter $\alpha$, called \textit{the strength of the coupling constant\rq s variation}, plays a substantial role, whereas in our model, all usual physical constants, except for $\Lambda$, are assumed to be nonvariant. We adopt this assumption because it is the simplest one. Therefore, the question of whether physical constants are unchanging or not remains unanswered. Qualitatively, much larger value of $\Lambda_{(dm)}$ in our model of the very early universe and much larger value of gravitational constant $G$ in Gupta's model would have a similar impact on the formation of early galaxies. The most important characteristics of our model is the possibility of the universe expansion in fifth dimensional direction, which reveals that our universe as a $4D$ Riemannian manifold is not a closed entity but is open to a higher dimensional realm.

\section*{Acknowledgments}
All of the authors express their sincere appreciation for the distinguished leadership of M. Ohtsu, Prof. Emer. of the University of Tokyo, who has been and is still driving unique \lq\lq off-shell science 
project'' originated from his lifelong dressed photon research in the field of nanophotonics.

\appendix

\section{Greenberg-Robinson theorem}
The essence of the theorem is that off-shell fields
is never compatible with the on-shell condition characterizing (generalized) free fields
in the special relativistic situation.
To precisely state the theorem, we briefly present the Wightman system of axioms.
In this appendix, we use the natural system of units, the unit system such that $c=1$ and $\hbar=1$.

The following family of four axioms is referred to as the Wightman system of axioms:
\begin{description}
\item[\text{[1. Field operator]}]
For a neutral field,
a field operator $\phi(x)$, $x\in\mathbb{R}^4$, is defined by an operator
on a Hilbert space $\mathcal{H}$.
\item[\text{[2. The covariance condition]}]
A (projective) unitary representation ($U$) of the Poincar\'{e} group exists,
$\mathcal{P}_+^\uparrow= \mathbb{R}^4\rtimes\mathcal{L}_+^\uparrow$ on $\mathcal{H}$
such that
\begin{equation}
U(a,L)\phi(x)U(a,L)^\ast=\phi(Lx+a)
\end{equation}
for all $(a,L)\in\mathcal{P}_+^\uparrow= \mathbb{R}^4\rtimes\mathcal{L}_+^\uparrow $.
\item[\text{[3. The causality condition]}]
$\phi(x)$, $x\in\mathbb{R}^4$, satisfy the local commutativity, that is,
\begin{equation}
[\phi(x),\phi(y)]=\phi(x)\phi(y)-\phi(y)\phi(x)=0
\end{equation}
for all pairs $(x,y)$ of mutually space-like points.
\item[\text{[4. Vacuum state and the spectrum condition]}]
A unit vector, $\Omega$ of $\mathcal{H}$, called the vacuum vector
and a spectral measure ($E$) of $\mathbb{R}^4$ on $\mathcal{H}$  exist, satisfying the following two conditions:
\begin{enumerate}
    \item[4.1] $\Omega$ satisfies $U(a,L)\Omega=\Omega$ for all $(a,L)\in\mathcal{P}_+^\uparrow$.
    $E$ satisfies
    \begin{equation}
        U(a,I)=\int_{\overline{V_+}} e^{ia\cdot p}\;dE(p)
    \end{equation}
for all $a\in\mathbb{R}^4$,
where $\overline{V_+}=\{x=(x^\mu)\in\mathbb{R}^4\,|\,x^2=x_\mu x^\mu\geq 0\;\text{and}\;x^0\geq 0\}$ and $a\cdot p=a^\mu p_\mu$
($\Omega$ then satisfies $E(\{0\})\Omega=\Omega$).
    \item[4.2] $\Omega$ is cyclic for $\mathcal{P}(\mathbb{R}^4)$, 
   the polynomial algebra over $\mathbb{C}$ generated by $\{\phi(x)\,|\,x\in\mathbb{R}^4\}$, i.e.,
    $\mathcal{H}=\overline{\mathcal{P}(\mathbb{R}^4)\Omega}=\overline{\{X\Omega\,|\,X\in\mathcal{P}(\mathbb{R}^4)\}}$.
\end{enumerate}
\end{description}
Strictly, every field operator is defined as an operator-valued distribution. However,
we omit the detail (see \cite{SW00} for example).
Moreover, no field operator is defined at each point $x$ of the Minkowski space, so that $\phi(x)$ is simply a symbolic notation.

A neutral field $\phi(x)$ is \textit{irreducible} if it satisfies the following condition:
If $B$ is a bounded linear operator on $\mathcal{H}$
such that
\begin{equation}
\langle \Omega|B\phi(x_1)\phi(x_2)\cdots \phi(x_m)\Omega\rangle
=\langle \phi(x_m)^\ast\phi(x_{m-1})^\ast\cdots \phi(x_1)^\ast\Omega|B\Omega\rangle
\end{equation}
for every natural number $m$ and points $x_1,\cdots,x_m$ of the Minkowski space,
then it is of the form $B=kI$, where $k\in\mathbb{C}$ and $I$ is the identity operator on $\mathcal{H}$.

The Fourier transform $\hat{\phi}(p)$ of a field operator $\phi(x)$ is defined by
\begin{equation}
\hat{\phi}(p)=\dfrac{1}{(2\pi)^2}\int_{\mathbb{R}^4} e^{-ip\cdot x}\phi(x)\;d^4x.
\end{equation}
A field $\phi(x)$ is called a \textit{generalized free field} if it satisfies the commutation relation
\begin{equation}
[\phi(x),\phi(y)]=iP(x-y)I
\end{equation}
for all $x,y\in\mathbb{R}^4$.
The function $P(x)$ is called the function defined by
\begin{equation}
P(x) = \int_0^\infty D_{\sqrt{\xi}}(x)\;d\sigma(\xi),
\end{equation}
where $\sigma$ is a measure on $\mathbb{R}_{\geq 0}=\{x\in\mathbb{R}\,|\,x\geq 0\}$,
\begin{equation}
D_{m}(x) =\dfrac{1}{(2\pi)^3}\int_{\mathbb{R}^3}
\sin (p^0 x^0) e^{-i {\bm p}\cdot{\bm x}} \dfrac{d{\bm p}}{p^0}
\end{equation}
and $p^0=\sqrt{{\bm p}^2+m^2}$.
The free Klein-Gordon field $\varphi(x)$ with square mass $m^2(>0)$, a typical example, satisfies
the commutation relation $[\varphi(x),\varphi(y)]=iD_m(x-y)I$ for all $x,y\in\mathbb{R}^4$.

\begin{theorem}[Greenberg-Robinson \text{\cite{Greenberg62, Robinson62}}]
Let $\phi(x)$ be an irreducible neutral field.
If an open subset exists, $O$, of $\mathbb{R}^4\backslash \overline{V_{\pm}}$ such that $\hat{\phi}(p) = 0$ for all $p\in O$, where $\overline{V_{\pm}} = \{x\in\mathbb{R}^4\,|\,x^2 = x_\mu x^\mu\geq 0\}$,
then $\phi(x)$ is a generalized free field.
\end{theorem}
The contraposition of this theorem implies that
\begin{center}
\textit{If an irreducible neutral field $\phi(x)$ is not a generalized free field, 
it holds that $\hat{\phi}(x)\neq 0$ for all $p\in \mathbb{R}^4$.}
\end{center}
This fact indicates that the involvement of off-shell momenta is specific to interacting quantum fields
and is never observed in (generalized) free fields at the four-dimensional Minkowski space.





\begin{thebibliography}{99}
\bibitem{dirac1} P. A. M. Dirac, Letters to the Editor, {\it NATURE}
{\bf 139} (1937), 323.

\bibitem{dirac2} P. A. M. Dirac, Long Range Forces and Broken Symmetries, {\it Proceedings of the Royal Society, London A}
{\bf 333} (1973), 403--418.

\bibitem{milne} E. A. Milne, {\it Kinematic Relativity} (Oxford, 1948).

\bibitem{canuto} V. Canuto and J. Lodenquai, Dirac cosmology, {\it Astrophys. J.} {\bf 211} (1977), 342--356.

\bibitem{fran} T. C. Van Flandern, Is the gravitational constant changing? {\it International Astronomical Union Colloquium} {\bf 63} (1981), 207--208.

\bibitem{reev} H. Reeves, On the origin of the light elements ($Z<6$), {\it Rev. Mod. Phys.} {\bf 66}
(1994), 193.

\bibitem{peeb} P. J. E. Peebles and B. Ratra, Cosmology with a Time-Variable Cosmological \lq\lq Constant\rq\rq, {\it Astrophysical J. Lett.} {\bf 325} (1988), L17. 

\bibitem{troit} V. S. Troitskii, Physical Constants and Evolution of the Universe, {\it Astrophysics and Space Science} {\bf 139} Issue 2, (1987), 389--411.

\bibitem{petit1} J. P. Petit, AN INTERPRETATION OF COSMOLOGICAL MODEL WITH VARIABLE LIGHT VELOCITY, {\it Mod. Phys. Lett. A} {\bf 3}, No.16, (1988), 1527-1532.

\bibitem{petit2} J. P. Petit and M. Viton, GAUGE COSMOLOGICAL MODEL WITH VARIABLE LIGHT VELOCITY: III. COMPARISON WITH QSO OBSERVATIONAL DATA, {\it Mod. Phys. Lett. A} {\bf 4}, No. 23, (1989), 2201--2210.


\bibitem{rand} Lisa Randall and Raman Sundrum, Large Mass Hierarchy from a Small Extra Dimension,
{\it Phys. Rev. Lett.} {\bf 83} (1999), 3370.

\bibitem{jost}
R. Jost, \textit{The General Theory of Quantized Fields};  (American Mathematical Society: Providence, RI, USA, 1963.)

\bibitem{antonio}
G. F. Dell\rq Antonio, Support of a field in $p$ space. {\it J. Math. Phys.} {\bf 2} (1961), 759--766.

\bibitem{ojima1}
I. Ojima, A unified scheme for generalized sectors based on selection criteria-order parameters of symmetries and of thermal situations and physical meanings of classifying categorical adjunctions. {\it Open Syst. Inf. Dyn.} {\bf 10} (2003), 235--279.

\bibitem{ojima2}
I. Ojima, Micro-Macro duality and emergence of macroscopic levels. {\it Quantum Probab. White Noise Anal.} {\bf 21} (2008), 217--228.

\bibitem{sakuma0}
H. Sakuma, I. Ojima, M. Ohtsu, Gauge symmetry breaking and emergence of Clebsch-dual electromagnetic field as a model of dressed photons. {\it Appl. Phys. A} {\bf 123} (2017), 750.

\bibitem{sakuma1}
H. Sakuma, I. Ojima, M. Ohtsu, and H. Ochiai, Off-Shell Quantum Fields to Connect Dressed Photons with Cosmology, {\it Symmetry} {\bf 12(8)} (2020), 1244; \\
doi:10.3390/sym12081244.

\bibitem{sakuma2}
H. Sakuma, Virtual Photon Model by Spatio-Temporal Vortex Dynamics. In {\it Progress in Nanophotonics} {\bf Vol. 5}, pp. 53--77, T. Yatsui, Ed., (Springer Nature, Switzerland, 2018).

\bibitem{mack}
G. W. Mackey, A theorem of Stone and von Neumann. {\it Duke Math. J.} {\bf 16} (1949), 313--326.

\bibitem{dhr1} S. Doplicher, R. Haag and J. E. Roberts, {\it Comm. Math. Phys.} {\bf 13} (1969), 1--23; {\bf 15} (1969), 173--200; {\bf 23} (1971), 199--230 \& {\bf 35} (1974), 49--85.

\bibitem{dr1} S. Doplicher and J. E. Roberts, {\it Comm. Math. Phys.} {\bf 131} (1990), 51--107; {\it Ann. Math.} {\bf 130} (1989), 75--199 (1989); {\it Inventiones Math} {\bf 98} (1989), 157--218.

\bibitem{sakuma5} Sakuma, H., Ojima, I. and Ohtsu, M., Perspective on an Emerging Frontier of Nanoscience Opened up by Dressed Photon Studies, {Nanoarchitectonics} {\bf Vol. 5} Issue 1, (2024), 1 -- 23.

\bibitem{sakuma6} H. Sakuma, I. Ojima, and M. Ohtsu, Dressed photon in a new paradigm of off-shell quantum fields, {Progress in Quantum Electronics} {\bf Vol. 55}, (2017), 74 -- 87.

\bibitem{clebsch} Lamb, S. H. {\it Hydrodynamics}, 6th ed. Cambridge University Press: Cambridge, UK, (1930), 248--249. 

\bibitem{sakuma3} H. Sakuma and I. Ojima, On the Dressed Photon Constant and Its Implication for a Novel Perspective on Cosmology, {\it Symmetry} {\bf 13}, issue 4, (2021), p. 593; \\ 
https://doi.org/10.3390/sym13040593.

\bibitem{sakuma4} H. Sakuma, I. Ojima, H. Saigo and K. Okamura, Conserved relativistic Ertel\rq s current generating the vortical and thermodynamic aspects of space-time, {\it Int. J. Mod. Phys. A} {\bf 37} No. 22, 2250155 (2022),
https://doi.org/10.1142/S0217751X2250155X.

\bibitem{mald} Maldacena, J. The large N limit of superconformal field theories and supergravity. {\it Adv. Theor. Math. Phys.} (1998), 2, 231-252.

\bibitem{penro1}
R. Penrose, Before the Big Bang: an outrageous new perspective and its implications for particle physics, {\it Proceedings of EPAC} (2006), \\
https://accelconf.web.cern.ch/e06/PAPERS/THESPA01.PDF 

\bibitem{penro2}
R. Penrose, Singularities and Time-Asymmetry, {\it General Relativity: An Einstein Centenary Survey}
(Cambridge Univ. Press,) (1979), pp. 581-638.

\bibitem{cart}
S. Helgason, {Differential Geometry, Lie Groups, and Symmetric Spaces} (Academic press, New York) (1978).

\bibitem{dona}
S. K. Donaldson, An application of gauge theory to four-dimensional topology, {\it Journal of Differential Geometry}, {\bf 18} (2), (1983), 279-315.

\bibitem{ojima3}
I. Ojima, Nakanishi-Lautrup B-Field, Crossed Product \& Duality, {\it RIMS Kokyuroku} {\bf 1524} (2006), 29--37.

\bibitem{hss} 
H. S. Snyder, Quantized space-time, {\it Phys. Rev.} {\bf 71}, (1947), 38.

\bibitem{liu}
Liu, H. \{Available online:\} What-is-the-best-estimate-of-the-cosmological-constant.
 \{https://www.quora.com\} ({accessed on 1 January, 2021}).

\bibitem{ert} H. Ertel, {\it Meteorol. Z.} {\bf 59} (9), (1942), 277-281.

\bibitem{aoki2} S. Aoki, T. Onogi and S. Yokoyama, Charge conservation, entropy current and gravitation, {\it Int. J. Mod. Phys. A} {36}, No. 29 2150201 (2021).

\bibitem{Landau}
Landau, L.D.; Lifshitz, E.M. \textit{Course of Theoretical Physics}, 2nd ed.; Volume 6 Fluid Mechanics; Elsevier: Oxford, UK, (1987).

\bibitem{witten} E. Witten, STRING THEORY DYNAMICS IN VARIOUS DIMENSIONS, \\ 
{arXiv:hep-th/9503124v2} (1995).

\bibitem{tom} M. Takesaki, {\it Lecture Notes Math.} {\bf 128} (Springer 1970). \\
doi:10.1007/BFb0065832, ISBN 978-3-540-04917-3

\bibitem{gupta} R. P. Gupta, \textit{JWST} early Universe observations and $\Lambda$CDM cosmology, {\it
Monthly Notices of the Royal Astronomical Society}, Vol. 524, Issue 3, (2023), 3385-3395. \\
https://doi.org/10.1093/mnras/stad2023

\bibitem{SW00} R.F. Streater and A.S. Wightman, \textit{PCT, Spin and Statistics, and All That}, (Princeton Univ. Press, 2000).

\bibitem{Greenberg62}
O.W. Greenberg, Heisenberg fields which vanish on domains of momentum space, J. Math. Phys. \textbf{3} (1962), 859--866.

\bibitem{Robinson62} D.W. Robinson, Support of a field in momentum space, Helv. Phys. Acta \textbf{35} (1962), 403--413.


\end{thebibliography}
\end{document}